\documentclass[12pt]{article}
\font\openface=msbm10 at10pt 

\def\R{{\cal R}}
\def\Z{{\cal Z}}
\def\H{{\cal H}}
\def\NaturalNumbers{{\hbox{\openface N}}}
\def\Integers      {{\hbox{\openface Z}}}
\def\braces#1{ \{ #1 \} }

\def\Had{\,{\scriptstyle \sharp}\,}

\def\journaldata#1#2#3#4{{\it #1} {\bf #2:} #3 (#4)}
\def\eprint#1{$\langle$#1\hbox{$\rangle$}}

\newcommand{\be}{\begin{equation}}
\newcommand{\en}{\end{equation}}
\newcommand{\bea}{\begin{eqnarray}}
\newcommand{\ena}{\end{eqnarray}}

\newcommand{\e}{\mbox{e}}

\newtheorem{Def}{Definition}
\newtheorem{Theo}{Theorem}
\newtheorem{Corol}{Corollary}
\newtheorem{Lemma}{Lemma}

\author{
Xavier Martin$^{a}$, Denjoe O'Connor$^{a,b}$ and R.D. Sorkin$^{c}$\\
\\ 
{\small\it $^{a}$ School of Theoretical Physics, DIAS,
10 Burlington Road, Dublin 4} \\
 \\ 
{\small\it $^{b}$ Depto de F\'{\i}sica, Cinvestav,}
{\small\it Apartado Postal 70-543, M\'{e}xico D.F. 0730.}\\
\\ 
{\small\it $^{c}$ Perimeter Institute, Waterloo, Ontario N2J 2W9} \\
{\small\it and }
{\small\it Dept. of Physics, Syracuse University,}
{\small\it Syracuse, New-York 13244-1130}
}

\title{The Random Walk in Generalized Quantum Theory}

\date{\today}

\begin{document}
\maketitle 

{\abstract
One can view quantum mechanics as a generalization of classical
probability theory that provides for pairwise interference among
alternatives.  
Adopting this perspective, we ``quantize'' the
classical random walk 
by finding, subject to a certain condition of ``strong positivity'',
the most general Markovian, translationally invariant ``decoherence
functional'' with nearest neighbor transitions.}

\section{Introduction}
In the causal set approach to quantum gravity \cite{SpaceCS} the basic
entity is taken to be a discrete causal order.  This, to our knowledge,
is the only candidate theory of quantum gravity, that naturally
predicts a cosmological constant of the right magnitude. 
(The predicted cosmological constant is related inversely to the square
root of the total number of spacetime elements in the universe, there
being roughly one element per Planck volume, \cite{lambdapred}.)
On a classical level, one has a well defined stochastic dynamical scheme
for causets, and much is understood about this ``CSG'' model,\footnote
{The initials stand for ``classical sequential growth''.}
especially about its special case of ``transitive percolation'' 
\cite{rideout,mors,avner,anpa,observables}.  
Although interesting and instructive in its own right, this model was
intended first of all as a stepping stone to quantum gravity, which in
this context would be a quantum dynamics related to the CSG model
analogously to how the quantum dynamics of a free particle is related to
the classical stochastic process of diffusion (``Wiener process'').
However, given the finitary nature of the causal set, it seems unlikely
that such a quantum process could be unitary, and it is natural to look
for a wider setting in which a suitable dynamics might be found.

Quantum measure theory (see below) offers such a setting, and our
primary goal in the present paper is to explore the analog of the
classical random walk in this setting.  More specifically, we will be
interested in Markov processes with nearest neighbor transitions.  This
problem seems particularly apt as a ``warm up'' for causal sets,
because, here also, the discreteness of the random walk obstructs a
unitary ``quantization'' of the classical stochastic process.\footnote 
{More precisely, no nontrivial unitary process exists which is Markovian
 and for which the random walker carries no internal variables and makes
 only nearest neighbor transitions \cite{dam}.  By allowing a one-step memory, or
 equivalently, endowing the walker with a two-valued internal state, one
 can salvage unitarity, and several models of this sort have been
 discussed in the literature \cite{dam,bogh}.}
In this way, the discrete case differs markedly from the continuous one:
the diffusion equation in the continuum has a unitary analog in the
Schr{\"o}dinger equation, whereas the random walk on the integers has,
strictly speaking, no unitary analog at all.

A second reason for interest in the quantal\footnote%
{Quantal will be used to refer to the generalised quantum measure
 theories studied in this paper, of which unitary quantum mechanics is
 only one particular case.}
random walk comes not from quantum gravity at its fundamental level, but
from a recent phenomenological theory which attempts to implement in a
concrete model the aforementioned suggestion from causal set theory that
the cosmological constant $\Lambda$ would be subject to $1/\sqrt{N}$
fluctuations.  This phenomenological theory \cite{lambdamodel}
introduces a (classical) random walk to act as the source of the
fluctuations, but it would seem more logical for the source to be
quantum in nature, since the fluctuations are supposed to be rooted in
an analog of the time-energy uncertainty principle.  In other words, it
would be natural to use a quantum random walk as the source of the
fluctuations, but first one should know what a quantum random walk is!

Thirdly, we believe that, independent of any application to quantum
gravity, the problem of the quantum random walk is of interest in
itself, as perhaps the simplest example of a situation where one is
forced to the type of generalization of unitary quantum mechanics called
``generalized quantum mechanics'' in \cite{gen} and
``quantum measure theory'' in \cite{Sorqmt,Sorqmti}.
Moreover, natural questions arise in this connection which seem not to
have been addressed so far, for example the question whether there
exists a quantum analog of the central limit theorem, which classically
almost guarantees the existence of a unique continuum limit.
If there does exist such a limit, one would expect it to be the
``Schr{\"o}dinger process'', i.e. the ``quantum stochastic process''
corresponding to the Schr{\"o}dinger equation.  
But whether or not some sort of central limit theorem obtains, one can
still ask whether any quantum random walk has the Schr{\"o}dinger
process as its limit.
To our knowledge, this question remains open (cf. \cite{dam,bogh}), 
but
one might hope to settle it using the techniques we develop herein.
Indeed, we hope that one of the processes we construct below will have
the desired limit.
Among other things, this would illustrate how a non-unitary
discrete process can become unitary in a continuum limit (and this in
turn would be of great interest for quantum gravity).

As just intimated, we will seek a theory of the quantal random walk 
formulated as a generalized measure theory \cite{Sorqmt,Sorqmti,gen,rob,
Chryssomalakos:2003ms}. 
When thought of in this way, 
quantum mechanics
is seen to reside in level 2 of an infinite hierarchy of generalized
measures, graded by the degree of interference they allow for (with
pairwise interference being characteristic of the quantum level). 
Let us briefly review this framework.

Classical probability theory assigns a probability $P(A)$ to 
a subset $A$ of the set of all ``configurations'',\footnote
{In general the configurations can be anything, but in the context of
 this paper they will comprise the possible trajectories of a random
 walker.}
and such probabilities satisfy the familiar Kolmogorov sum rule:
\begin{equation}
   P(A \sqcup B) = P(A) + P(B)
\end{equation}
where $A \sqcup B$ is the disjoint union of $A$ and $B$.
This sum rule yields measure-theory in the classical
sense, and, physically, it is appropriate for the description of 
classical stochastic processes such as Brownian motion.  
But the
additivity of classical probabilities is only the first, and most
restrictive, in a hierarchy of possible sum-rules, each of which
implies its successor.  

The next weaker sum-rule defines a {\it generalized measure theory}
which includes quantum mechanics as a special case.  We will refer to a
generalized measure satisfying this sum-rule as a ``quantum'' or
``quantal'' measure and a physical process described by such a measure
as a quantal process or a generalized quantum process.  The fact that
quantum probabilities can be expressed ``as the squares of amplitudes''
is a natural consequence of this sum rule.  Even weaker sum-rules
provide further generalizations of classical probability theory that we
will now define, but without then considering them any further.

In its quantal measure formulation, the familiar quantum mechanics of
point-particles does its job by furnishing ``generalized probabilities''
for sets of trajectories.
More formally, 
it associates to a set $A$ of trajectories 
a non-negative real number denoted $\mu(A)$ or $|A|$
which is called its {\it quantal measure}; 
and it is this measure that enters into the sum-rules. 
The deviation from the classical case consists in the fact that
the {\it interference term}
$$
   I(A,B) := |A \sqcup B| - |A| - |B|
$$
between two disjoint sets of trajectories $A$ and $B$ is in general
not zero.   

Consider the following generalizations of this interference term.
\bea
  I_1(A) & \equiv & |A| \\
  I_2(A,B) & \equiv & |A \sqcup B| -|A| -|B| \\
  I_3(A,B,C) & \equiv & |A \sqcup B \sqcup C| -|A \sqcup B| - |B \sqcup
   C| -|A \sqcup C|+ |A| +|B| + |C|, \end{eqnarray}
or in general,
\be
   I_n(A_1,A_2,\cdots,A_n) \equiv |A_1 \sqcup A_2 \sqcup \cdots A_n|
   - \sum | \bigsqcup_{j=1}^{n-1} A_{\sigma_j} | 
   + \sum | \bigsqcup_{j=1}^{n-2} A_{\sigma_j} | 
   - \cdots \pm \sum\limits_{j=1}^{n} |A_j| , \en
where all the sets $A_i$ 
are mutually disjoint
and $\sigma$ 
indexes the subsets of $\braces{1,2,\cdots,n}$ with the appropriate cardinality.

These expressions are related serially by the identity
\begin{equation} 
  I_{n+1}(A_0, A_1, A_2, \cdots, A_n) =
                I_{n}(A_0 \sqcup A_1, A_2, \cdots, A_n)
                -  I_{n}(A_0, A_2, \cdots, A_n)
                -  I_{n}(A_1, A_2, \cdots, A_n)
\end{equation}
So, for each $n$ one obtains a possible sum-rule by setting $I_n$ to
zero, and one sees that if the $n^{th}$ such sum-rule is satisfied, then
the $(n+1)^{st}$ is automatically also satisfied. 
Hence the sum-rules form a hierarchy of ever decreasing strength. 
The first sum-rule in the hierarchy, $I_1 = 0$, trivializes the measure
and is therefore uninteresting.
The second expresses precisely the additivity of classical measure
theory, or equivalently the additivity of classical probabilities, when
they are regarded as set-functions in the Kolmogorov manner.
Accordingly, the third sum-rule, $I_3(A,B,C)\equiv 0$, defines 
a generalization of measure theory which,
although it allows for pairwise interference of alternatives, 
still preserves most
of the additivity of classical probabilities.  
This is the level of quantal measure theory.\footnote%
{In addition to this sum rule, what is called quantum measure theory in
 \cite{Sorqmti} is characterized by the axiom that no set of measure
 zero can interfere with any other (disjoint) set: $|A{\sqcup}N|=|A|$
 for all $N$ such that $|N|=0$.  This will follow trivially from
 ``strong positivity'', as shown in Theorem \ref{prodineq}.}
The fourth and higher sum-rules define still more general forms of
measure theory which may be regarded as natural extensions of quantum
mechanics, and which incoroporate interference among successively larger
numbers of alternatives. 

Plainly, the interference term is totally symmetric in its arguments,
and thus 
$I_{n+1}$ vanishes if and only if $I_n$ is ``additive'' in
each argument, given 
their mutual disjointness.
Thus each sum-rule is associated with a kind of multilinearity (or
rather multi-additivity) of the function which measures the failure of
the next stronger sum-rule to hold.  At the quantum level,
specifically, we learn that $I_2$ is bi-additive, and the peculiar
quadratic relationship between quantum amplitudes and probabilities
corresponds directly to this feature of $I_2$.  In the current work we
limit ourselves to the quantal case as defined by the $n=3$ sum-rule,
\begin{equation}
    |A \sqcup B \sqcup C| -  |A \sqcup B| - |B \sqcup C| -|A \sqcup C|
     + |A| +|B| + |C| = 0.           
\label{quantum_sum_rule}
\end{equation}
Given this sum-rule, we have that $I_2$, which we will subsequently
denote by $I$ is bi-additive in the sense that 
\begin{equation}
    I(A \sqcup B, C) = I(A,C) + I(B,C),  
\end{equation}
whenever $A$, $B$ and $C$ are mutually disjoint.  
Taking
\begin{equation}
     I(X,X) \equiv 2 \, |X|.
\end{equation} 
as the value of $I$ on equal arguments, we can extend the above
definitions to arbitrary arguments, and in fact the value of $I$ is
completely determined by bi-additivity and can be given in terms of
the quantal measure in, for example, either of the
following equivalent forms \cite{Sorqmt}:
\begin{eqnarray}
 I(A,B) & = & |A\cup B|+|A\cap B|-|A\backslash B|-|B\backslash A|,
 \nonumber\\ \label{interference_functional} \\
 I(A,B) & = & |A\Delta B|+|A|+|B|- 2\, |A\backslash B|-2\, |B\backslash
 A|.\nonumber 
\end{eqnarray}
In these equations the symbol `$\backslash$' denotes set-theoretic difference
and `$\Delta$' denotes symmetric difference.

Note that any generalized measure obeying the quantum sum rule $I_3=0$
can be expressed in the form $|X|=I(X,X)/2$, where $I$ is the
bi-additive, real-valued set function of
(\ref{interference_functional}).  Conversely, we could begin with such
a set-function whose diagonal values are all non-negative, and use it
to define a quantal measure $|\cdot|$ obeying the sum rule
(\ref{quantum_sum_rule}). 

In unitary quantum mechanics 
(pretending that the set of all possible particle
paths has finite cardinality) 
the measure of any set 
$A=\braces{x,y,\cdots,z}$ 
of paths 
can be expressed formally as
$|A| = ``(1/2) I(x+y+\cdots z,x+y+\cdots z)$'',
which is to be evaluated by expanding out the sums via bilinearity and
interpreting $I(x,y)$ as $I(\braces{x},\braces{y})$ with 
\begin{equation}
  I(\braces{x},\braces{y}) = \delta(x(T),y(T)) \; e^{i S(x)} e^{-i S(y)} +
  (\mbox{complex conjugate}),
\end{equation}
where $T$ is a ``truncation time'' lying to the future of the properties
defining $A$, and $x$ and $y$ are truncated paths with final end points at
time $T$ (see \cite{Sorkin_role_oftime} for more details).  
A striking feature of unitary quantum mechanics from this point of view
is the presence in $I(x,y)$ of a delta-function which ``ties together''
the endpoints created by truncating the paths.

A glance at this last equation reveals that $I(A,B)$ is twice the real
part of the so called {\it decoherence functional} $D(A,B)$ where
\begin{equation}
    D(x,y) = \delta(x(T),y(T)) \; e^{i S(x)} e^{-i S(y)}.
\label{qmD}
\end{equation}
Notice that $D(x,y)$ exhibits the symmetry property of a hermitian matrix:
\be
      D(y,x)=D^*(x,y).
\en 
For our purposes,
the decoherence functional can be taken to be the
more primitive object,
since the Markov property we wish to implement is naturally expressed
in terms of $D$ but not $I$ or the generalized measure $\mu$ itself.
A decoherence functional specifying a quantal
measure (i.e. a level two generalized measure) will be required
to satisfy:  
\begin{itemize}
  \item[{\it i)}] bi-additivity, 
  \be D(A\sqcup B,C)=D(A,C)+D(B,C) 
   \hbox{ \quad and\quad } D(A,B\sqcup C)=D(A,B)+D(A,C),\label{bi-add}
  \en
  \item[{\it ii)}] hermiticity \be D(A,B)=D^*(B,A),\label{symm} \en
  \item[{\it iii)}] and positivity \be |A|=D(A,A)\geq0.\label{posi} \en
\end{itemize} 

There are two simple examples which can be derived from a classical
probability measure $p$. The first is given by the classical
(non-interfering) interference function 
\be 
D(A,B)= p(A\cap B)  \ .
\label{Ex1}
\en 
The second example has no immediate interpretation and is given by
\be
 D(A,B)=p^*(A)\, p(B).
\label{Ex2} 
\en 
Both these decoherence functionals satisfy {\it i)}, {\it ii)} and {\it iii)} 
above.  (Obviously, the classical probability measure is real, and
thus there was no need to include a complex conjugation in the second
example.  However, if we do so, then the expression remains valid in the
more general case where $p$ is an arbitrary, additive {\em complex}
function of its argument.) 

In this article we will focus on the decoherence functional $D$ and on
paths whose positions are restricted to a discrete lattice in both space
and time.  We will seek the most general translationally invariant
processes with nearest neighbor transition amplitudes that form level
two generalized measures, i.e.  that yield a $D$ satisfying {\it i)},
{\it ii)} and {\it iii)} above.  The choice of nearest neighbor
transition amplitudes is made for simplicity. The construction
presented here could easily be generalised to processes with transitions
to any finite number of neighbors.

Our quantal processes are analogues of random walks based on coin
tossing in classical probability theory.  The additional complications
we encounter in comparison with the classical case are due mainly to two
differences: the more involved nature of the positivity condition, and
the possibility that the two walkers corresponding to the two arguments
of $D$ can ``interact''.  This ``interaction'' generalizes the delta
function endpoint interaction which characterizes the standard quantum
case with its unitary evolution.  The new processes go beyond quantum
mechanics in that the evolution is not unitary but satisfies a
weaker\footnote
{Even though positivity is strictly weaker than unitarity, it is {\it not}
 the case that the type of process we will construct contains unitary
 evolution as a special case.  We discuss this subtlety further below
 in Section 3.0.}
requirement expressing the positivity of the quantal measure.  One might
think that dropping unitarity allows significant extra freedom,
nevertheless, we find the requirements of a well defined level two
measure to be remarkably restrictive.

The processes we find are quantal in the sense that they involve
transition amplitudes with only pairwise interference.  
They are also nearest neighbor in that 
the walker can take at most one step to the left or right,
and
we are optimistic that an appropriate limit of them will be unitary.
Because of their
simplicity and closeness to quantum processes of the more traditional
sort, we expect that they will yield additional insights into quantum
mechanics itself.  We hope as well that ideas in this work will help
lead us to a physically suitable quantum dynamics for causal sets.

In the next section, we consider further the properties that the
decoherence functional $D$ must possess in order to yield a well defined
quantal measure, and we introduce to this end a strengthened form of the
positivity condition, a condition we call ``strong positivity''.  We also
introduce the transfer matrix that defines the Markovian processes we
are interested in.  In section \ref{transferM} we find the most general
translationally invariant and positive transfer matrix corresponding to
a random walk with nearest neighbor transitions.  In section \ref{pptm}
we parameterize the general solution to the problem before concluding in
section \ref{conclusion}.

\section{The decoherence functional and strong positivity}
First a few definitions. 
The ``completed history'' of the walker's motion for $0\le{t}<\infty$
will be called a {\it trajectory} or {\it path}. 
We will assume that the path is {\em originary}, i.e.  begins at the
origin, and at each step either stays put or moves one unit to the right
or left.
A trajectory is thus an infinite sequence of integers,
$(0,n_1,n_2,n_3,\cdots)$.  
It can also be resolved into a
sequence of {\it steps}, 
$\xi_0=(0,n_1)$, 
$\xi_1=(n_1,n_2)$, 
$\xi_2=(n_2,n_3)$, 
etc.
The set of all trajectories will be denoted $\Omega_\infty$, and the
decoherence functional will therefore have its domain in the set of all
sets of trajectories $2^{\Omega_\infty}$.
We will also need the notion of truncated trajectory or path, that is a
trajectory defined only from the initial time $t=0$ to some finite
``truncation time'' $T$. 
The set of all such truncated paths will be denoted by $\Omega_T$.
Clearly the truncated paths do not live in the same space as the
trajectories on which the decoherence functional itself is defined.
However, the set of truncated paths $\Omega_T$ can be embedded into
$2^{\Omega_\infty}$ through the notion of {\it cylinder set}. 

The cylinder set $Z(\gamma)$ associated with a truncated path $\gamma$
of length $T$ is the set of all trajectories which coincide with
$\gamma$ until time $T$.  It should be clear that this mapping $Z$ from
truncated paths $\bigsqcup_T \Omega_T$ to sets of trajectories
$2^{\Omega_\infty}$ is injective and therefore defines a bijection
between the space of all truncated paths and that of all cylinder sets,
denoted $\Z$.  By dint of this equivalence, truncated paths and cylinder
sets will normally be identified in the following, and reference to the
mapping $Z$ will almost always be omitted.  We also make the convention
that the empty set is a cylinder set.

Since we are interested in random walks without memory (``Markov
processes''), the decoherence functional $D$ can be defined in stages,
in a manner we describe in detail in the next subsection \ref{dfftm}.
In analogy with what is done for a classical Markov process, 
we will first define the decoherence functional inductively on the cylinder sets 
and then extend it by additivity to more general sets of paths.

Key to this extension are two ``algebraic'' properties of the cylinder sets: 
first that the intersection $A{\cap}B$ of any two cylinder sets is a
cylinder set (in fact $A$ and $B$ are either disjoint or nested);
and 
second that the complement of any cylinder set is a finite disjoint
union of cylinder sets.\footnote
{A system of sets with these properties is sometimes called a
 ``semiring'' of sets or an ``abstract interval family''.}
It is not difficult to verify these properties if one thinks of the
cylinder sets as truncated paths.  Notice in particular that a cylinder
set of length $T$ is the disjoint union of all the cylinder sets of
length $T+1$ that it contains:
\be 
    \gamma^{T}=\bigsqcup_{\sigma} \gamma^T \# \sigma, \label{consistency} 
\en
where $\sigma$ represents all the possible steps the path $\gamma^T$ of
length $T$ can take in passing from $T$ to $T+1$, and the sign $\#$
represents the addition of an extra step to a truncated path.  (By
assumption, there can be only a finite number of possibilities for
$\sigma$.)
Now let $\R$ be the family of all disjoint unions of a finite
number of cylinder sets, {\it i.e.} all subsets
$A\subseteq\Omega_\infty$ 
of the form 
$A=\sqcup_{i=1}^n\gamma_i$,
where the $\gamma_i$ are cylinder sets.  
Using the two properties enumerated above, one can prove 
without difficulty
that $\R$ is a
``set algebra'': 
it is closed under the operations of union, intersection and complementation.  
Once $D$ has been defined consistently on cylinder sets, 
it extends uniquely to $\R$ via bi-additivity (\ref{bi-add}),
and this extension is consistent by the same two properties.  
(Clearly hermiticity (\ref{symm}) will be preserved by the extension.)
Thus $D$ is naturally defined on $\R$.

Notice
that $D$ cannot be defined arbitrarily on cylinder sets because one
such set can be the disjoint union of a finite number of others, 
as in (\ref{consistency}),
and
then (\ref{bi-add}) entails a consistency condition.  In our case, we
will first define $D(\gamma_1,\gamma_2)$ for $\gamma_1$ and $\gamma_2$
of equal length, and consistency will then reduce to a simple condition
corresponding to (\ref{consistency}) that we will make explicit below.
The general expression for $D$ on the domain $\R$ will then be
\be  
   D\left(H=\bigsqcup_{i=1}^n \gamma_i ,
          K=\bigsqcup_{j=1}^{n'} \gamma'_j \right)
 =\sum_{ij} D(\gamma_i,\gamma'_j)  \ ,
 \label{e17}
\en
where the $\gamma$'s are cylinder sets which, by (\ref{consistency}) 
and (\ref{bi-add}), can all be taken to have the same length.

In our definition of $D$ on cylinder sets, the hermiticity property
(\ref{symm}) will hold by construction.  Its general validity will then
follow immediately from (\ref{e17}).
We thus satisfy by construction both bi-additivity and hermiticity.
The positivity condition (\ref{posi}) is a different story, however, and
its implementation will occupy us throughout this paper.

The positivity condition we will implement will be somewhat stronger
than (\ref{posi}), at least in appearance, and will be called {\it
strong positivity}.  Although possibly more restrictive, this condition
has proved easier to work with than (\ref{posi}), and we prefer it for
that reason.  In addition, it has an independent significance as a
condition whose satisfaction allows one to derive a Hilbert space from
the quantal measure/decoherence functional.  And this in turn might be
helpful in attempting to extend $D$ and $\mu$ to a larger domain of
definition than just $\R$.  Let us briefly consider these
interconnections.

On one hand, many questions of physical importance correspond, not to
sets in $\R$ but to ``limits'' of such sets.
For example, the set of all trajectories $\gamma\in\Omega_\infty$ that
eventually return to the origin is not a finite but a {\it countable}
union of cylinder sets.
Ideally, therefore, one would like to extend $D$ ``by continuity'' 
from $\R$ to the full $\sigma$-algebra generated by the cylinder sets.
On the other hand, the techniques which accomplish this in the case of a
classical measure do not go through in the quantum case, because
$D(A,B)$ is neither positive nor bounded.
A different technique of ``completion'' is therefore called for.
In attempting to meet this need, one might be able to take advantage of
the possibility that $\R$ can be made into a Hilbert space with the aid
of the decoherence functional $D$.  

The construction of the Hilbert space $\H$ proceeds in ``GNS'' 
(Gel'fand-Naimark-Segal)
fashion.
One interprets $D$ as an ``inner product on $\R$'' and obtains $\H$ as
the set of formal linear combinations of elements of $\R$.  More
exactly, one quotients this space by the subspace of null vectors and
then completes the quotient to arrive at $\H$.
If one does this in the unitary case, the resulting Hilbert space $\H$
is none other than the one normally employed as the state-space of the
quantum system in question.  Thus, one recovers the quantum Hilbert
space $\H$ directly from the decoherence functional $D$, without any
reference to canonical commutation relations, classical phase spaces, or
other auxiliary structures
\footnote
{Compare the analogous construction of \cite{Glimm:1981xz}.}.
Now, nothing limits this construction to the unitary case, but in order
for it to succeed in general, (\ref{posi}) must be strengthened to the
requirement that we will call strong positivity (and which is
automatically satisfied in the unitary case).  As already mentioned, we
will see that this requirement also emerges naturally in the attempt to
guarantee (\ref{posi}) for our quantal random walk.

\begin{Def} 
   A decoherence functional $D$ is {\bf strongly positive} if 
   it fulfills either (and therefore both) of 
   the following two equivalent conditions:
\begin{itemize}
\item[{\it i)}] for any finite collection of sets  
  $A_i\in\R$, $1\leq i\leq n$, the (hermitian) matrix
  $N_{ij}=D(A_i,A_j)$ is positive;
\item[{\it ii)}]
  the (finite hermitian) matrix
  $M_{\gamma_1,\gamma_2}=D(\gamma_1,\gamma_2)$,  $\gamma_i\in\Omega_T$  
  is a positive matrix for all
  $T$.\end{itemize}
\end{Def}
Note that, as the names suggest, strong positivity implies weak
positivity, as can be seen by simply taking $n=1$ in the first
definition of strong positivity.  
It should also be clear that the
first definition of strong positivity implies the second by choosing
the finite sets $A_i$ to be the cylinder sets $\gamma_i$ in $\Omega_T$. 
It remains only to prove the reverse implication.

Let us therefore suppose that the matrix $M_{\gamma_1,\gamma_2}$ is
positive and let $N_{ij}=D(A_i,A_j)$. 
Writing the $A_i$ as sets of
truncated trajectories $\gamma^T$ for $T$ sufficiently
large, and introducing a generic column vector $\psi$
with components $\psi_i$, we have
\bea
  \psi^\dag N\psi 
   & = & 
   \sum_{\gamma_1,\gamma_2 \in \Omega_T} M_{\gamma_1,
  \gamma_2} \left( \sum_{i,\gamma\in A_i} \psi_i \delta_{\gamma_1,\gamma}
  \right) ^* \left( \sum_{j,\gamma'\in A_j} \psi_j \delta_{\gamma_2,\gamma'}
  \right) \\
   & = & 
  \sum_{\gamma_1,\gamma_2 \in \Omega_T} M_{\gamma_1,\gamma_2} \rho^*(\gamma_1)
   \rho(\gamma_2) \geq 0,
\ena
where the positivity of $M$ was used and we defined
\be
  \rho(\gamma')=\sum_{i,\gamma\in A_i} \psi_i \delta_{\gamma',\gamma}  \ .
\en
This concludes the demonstration of the equivalence between the two
definitions. 

As already noted,
we have the following obvious theorem:
\begin{Theo} A strongly positive decoherence functional is positive. \end{Theo}
Indeed, for any set of truncated paths and $A=\sqcup \gamma_i$, 
\be
  \mu(A)=\sum_{ij} D(\gamma_i,\gamma_j)=v^*_{\gamma_1}M_{\gamma_1\gamma_2} 
   v_{\gamma_2} \geq 0,
\en
where the notation is that of the second definition {\it ii)}
and $v_{\gamma}$ is a vector 
made of $0$s and $1$s which is $1$ if $\gamma\in A$ and $0$
otherwise.  Thus, in contrast with strong positivity, we see that
plain positivity only requires that the quadratic form associated with the
hermitian matrix $M_{\gamma_1\gamma _2}$ be positive when applied to
column vectors made of $0$s and $1$s.
Of course, 
an even simpler proof is possible, if we start from the first form of
Definition 1:  simply apply it
to the single set $A$ and observe that the $1\times 1$ positive
matrix $N$ is simply $[\mu(A)]$.

Since strong positivity is associated with a Hilbert space structure,
some familiar properties arise from it, such as the following.
\begin{Theo} 
For any two sets $A$ and $B$ in $\R$, $|D(A,B)|^2 \leq |A|\, |B|$ .
\label{prodineq} 
\end{Theo}
This is just the statement that the determinant of the positive
hermitian $2\times 2$ matrix obtained from definition {\it i)} of
strong positivity and the collection of sets $(A,B)$ is positive. As
observed in footnote $5$, this theorem guarantees us a
quantum measure theory in the sense of \cite{avner}.

\subsection{Decoherence functional from a transfer matrix} \label{dfftm}
In the following, we will be interested in a process whose
decoherence functional on truncated paths is defined iteratively
through a transfer matrix, which will be denoted $a$.
Since this is in close analogy to the Markov property of classical
probability theory, one may call such a process ``Markovian''.

Before proceeding, it is convenient to expand on our notation. A generic
truncated trajectory will be written as $\gamma$.  If it has length $T$,
this will be indicated by a superscript $\gamma^T$.  The position of a
path $\gamma$ at a given time $t$ will be written as $\gamma(t)$.  Since
we are interested in a quantum random walk with nearest neighbor
transitions, at each step, a truncated trajectory will only be allowed
three possible moves: to the same position, or to the nearest left or
right one.  Its successive moves can thus be expressed by a finite time
sequence of integers $\sigma(t)\in\braces{-1,0,+1}$.  Since we restrict
our consideration to originary paths, i.e. paths $\gamma$ beginning at
the origin, a truncated trajectory is entirely defined by its associated
sequence $\sigma_\gamma(t)$,
\bea
  \gamma(t+1) & = & \gamma(t)+\sigma_\gamma(t)\\ \gamma(t) 
   & = & \sum_{t=0}^{t-1} \sigma_\gamma (t).
\ena

As represented already in (\ref{consistency}), a truncated path of
length $T$ can be extended to a truncated path of length $T+1$ by
defining its position at the next step $T+1$, which as explained above
is just given by an integer $\sigma\in \braces{-1,0,+1}$.  The sequence
of length $T+1$ which coincides with $\gamma^T$ for the first $T$ steps
and which has position $\gamma^T(T)+\sigma$ at time $T+1$ will be
denoted $\gamma^T\#\sigma$.  With these choices and definitions,
Eq. (\ref{consistency}) becomes simply
\be 
  \gamma^T
  = (\gamma^T\#-1)  \sqcup (\gamma^T\#0) \sqcup (\gamma^T\#1) \ . \label{thid}
\en 
Conversely, a truncated path $\gamma^T$ of length $T$ can be truncated
further to a length $T-1$ path $\gamma^{T-1}$, by taking the latter to
coincide with $\gamma^T$ on the first $T-1$ steps.

Now, a transfer matrix will give transition amplitudes from time $T$ to
time $T+1$, for the {\it pair} of truncated paths $(\gamma_1,\gamma_2)$. 
Since
we are considering a one step Markov process, this amplitude should only
depend on the positions 
of the truncated paths at times $T$ and $T+1$.
Furthermore, assuming spatio-temporal homogeneity, it should not depend
explicitly on the step $T$ considered, and it should depend only on the
relative position of the two paths, not on their absolute location.
Thus the transfer matrix can be written as
$a(n_1,\sigma_1,n_2,\sigma_2)$ where $n_i$ are integers and $a$ only
depends on $n_i$ through the difference $n_1-n_2$:
\be
  a(n_1,\sigma_1,n_2,\sigma_2)=a(n_2-n_1,\sigma_1,\sigma_2) \ .
\label{aparam}\en
In the rest of the paper, we will use for $a$ whichever notation is most
convenient at the time. 

With these notations, the  decoherence functional on cylinder sets is
given recursively by
\be
  D(\gamma_1^T\#\sigma_1,\gamma_2^T\#\sigma_2)
 = D(\gamma_1^T,\gamma_2^T) 
   a(\gamma_1^T(T),\sigma_1,\gamma_2^T(T),\sigma_2) \ .  
 \label{Ddef} 
\en
The hermiticity condition (\ref{symm}) then requires that 
\be 
 a(n_2,\sigma_2,n_1,\sigma_1)=a^*(n_1,\sigma_1,n_2,\sigma_2)\ .
\label{asym} 
\en 
As for the self-consistency condition (\ref{consistency}), it
constrains the transfer matrix to satisfy 
the equation,
\be 
  D(\gamma_1^T,\gamma_2^T) 
  \left( 
  \sum_{\sigma_1,\sigma_2} a(\gamma_1^T(T),\sigma_1,\gamma_2^T(T),\sigma_2)-1
  \right) =0
\label{aconstr} 
\en 
for any time $T$ and 
any pair of truncated paths $(\gamma_1^T,\gamma_2^T)$
and integers $(\sigma_1,\sigma_2)$. 
In general, there will always be at
least one time $T$ and pair of truncated paths $(\gamma_1^T,\gamma_2^T)$
ending at a given pair of lattice positions $(n_1,n_2)$ such that 
$D(\gamma_1^T,\gamma_2^T)\not=0$; 
and therefore our constraint simplifies to 
\be
 \sum_{\sigma_1,\sigma_2} a(n_1,\sigma_1,n_2,\sigma_2)=1\ . 
\label{Qconstr} 
\en

However, this is not always true, and 
there exist important special cases in which it fails.
One such is the classical random walk, for which 
there is no interference and $D$ assumes the degenerate form (\ref{Ex1}).
In this case 
(since distinct members of $\Omega_T$ are disjoint as cylinder sets),
$D(\gamma_1^T,\gamma_2^T)$ vanishes unless 
$\gamma_1^T=\gamma_2^T$,
and the transfer matrix is necessarily diagonal in
$(\sigma_1,\sigma_2)$.
The constraint (\ref{aconstr}) then reduces to 
\be
 \sum_{\sigma} a(n,\sigma,n,\sigma) = 1 \ .
\label{clasconstr} 
\en
In fact, if we write the transfer matrix in the form
$a(n_2-n_1,\sigma_1,\sigma_2)$, we see that it does not depend on $n$ at
all in this case, and (\ref{clasconstr}) reduces further to 
$\sum_{\sigma}a(\sigma)=1$, which is just the Markov sum rule for a
classical stochastic process.  
Notice that, here, the transfer matrix is strictly speaking undefined
unless $n_1=n_2$.

As we have just seen, the classical random walk is characterized by the
condition that the ``walker'' and ``antiwalker'' never separate.
Between this and the generic case, there may exist intermediate cases in
which the walker and antiwalker can move apart, but never by more than
some fixed distance $d$.  Such processes, if they exist, might offer an
interesting transition between the fully classical and fully quantum
situations.  

Obviously, the transfer matrix $a$ carries full information about the
decoherence functional $D$, and it would be useful to translate the
strong positivity condition on $D$ to a condition on $a$.  By doing so,
one obtains theorem 3 below.  In the statement of this theorem, $a$ is to
be regarded as a matrix with rows and columns labeled by the pairs
$\xi_1=(n_1,\sigma_1)$ and $\xi_2=(n_2,\sigma_2)$ respectively.  That
is, the rows and columns correspond to {\it steps} of the two respective
paths $\gamma_1$ and $\gamma_2$, not, as one might think, to their
successive {\it locations}, which would label the rows and columns by
successive values of the pair $(n_1,n_2)$. 
Notice that $a$ is an infinite matrix.  We will define such a matrix to
be {\it positive} if every finite principal submatrix is positive as a finite
matrix.  
Equivalently,
$\sum_{\xi\eta}a_{\xi\eta}\psi^*_\xi\psi_\eta\ge0$
for every ``column vector'' $\psi$ having only a finite number of
nonzero components $\psi_\xi$.

Notice also that the positivity condition on $D$ which figures in
Theorem 3 is not ``weak'' positivity (\ref{posi}) but strong positivity
as defined in the previous subsection.  We do not know whether there
exists any simple translation of weak positivity into a property of the
transfer matrix, and this is one reason why we have been led to work
with strong positivity instead.

\begin{Theo} \label{Th1}
  If the transfer matrix $a$ is positive, then the
  corresponding decoherence functional $D$ defined through (\ref{Ddef})
  is strongly positive. 
\end{Theo}

In order to prove this theorem we will build up the decoherence
functional as a kind of product involving the transfer matrix.
However, the product in question is not the ordinary matrix product,
but rather what is called the Hadamard product, denoted $\Had$ 
in the
following, which multiplies two matrices $M^{(1)}_{ij}$ and
$M^{(2)}_{ij}$ entry by entry:
\be
  (M^{(1)}\Had M^{(2)})_{ij}=M^{(1)}_{ij}M^{(2)}_{ij}.
\en
The proof of the theorem then hinges on the following
lemma.
\begin{Lemma} 
  The Hadamard product of two positive matrices is also a positive matrix.
\end{Lemma}

The proof is quite simple. 
Since $M^{(1)}$ is positive and therefore
hermitian, it can be put into the following diagonal form\footnote
{Because the matrices here are playing the role of quadratic forms, 
 the $\lambda$'s have no real significance, and could be absorbed
 into the $v$'s}
\be
    M^{(1)}=\sum_i \lambda_i v^{(i)} v^{(i)\dag},
\en 
where the $\lambda_i\geq 0$
are its eigenvalues and the $v^{(i)}$ are a corresponding basis of
orthonormal eigenvectors.  Then, given a generic vector $\psi$, 
\bea
  \psi^\dag (M^{(1)} \Had M^{(2)}) \psi 
  & = & 
  \sum_{ijk}
  \lambda_i (\psi_j v_j^{(i)})^* M^{(2)}_{jk}(\psi_k v_k^{(i)})\\ 
  & = & 
  \sum_i
  \lambda_i \rho^{(i)\dag} M^{(2)} \rho^{(i)} \geq 0,
\ena 
since each of the
terms in the sum is positive by definition of the positivity of
$M^{(2)}$, and where 
\be 
     \rho^{(i)}=\psi \Had v^{(i)}.
\en

{\it Proof of the Theorem:} 
The theorem is proved by checking that if $a$ is positive then the
matrices $D(\gamma_1,\gamma_2)$ for $\gamma_i\in\Omega_T$ are positive
for all $T$.  This, in turn, is proved by induction on $T$.  For
$T=0$, the result is trivial since then $\Omega_T$ contains only one
element, with measure one (as conventionally normalized).

Let us now suppose that the theorem is true at step $T$: $D$ is
positive on 
$\Omega_T$. Then we need to prove that the matrix 
\be
 D(\gamma_1^{T+1},\gamma_2^{T+1})=D(\gamma_1^T,\gamma_2^T) a(\gamma_1
 (T),\sigma_1(T),\gamma_2(T),\sigma_2(T))
\en
is a positive matrix. 
To use the Hadamard positivity lemma on the right hand side, 
both matrices $a$ and $D$ must be extended to matrices
of the size of $\Omega_{T+1}$,
which we will call $a^e$ and $D^e$. 
This is easily done by some appropriate truncations: 
\bea
 a^e(\gamma_1^{T+1},\gamma_2^{T+1}) & = & a(\gamma_1^{T+1}(T),\gamma_1 
 ^{T+1}(T+1)-\gamma_1^{T+1}(T),\gamma_2^{T+1}(T),\gamma_2^{T+1}(T+1)-
  \gamma_2^{T+1}(T)) \\
  D^e(\gamma_1^{T+1},\gamma_2^{T+1}) & = & D(\gamma_1^T,\gamma_2^T).
\ena 
Then, by the Lemma, it suffices to
prove that the two matrices $a^e$ and $D^e$ are positive. 

This is 
actually a simple matter since for any vector $\psi$ indexed on
$\Omega_{T+1}$, 
\bea 
  \psi^\dag a^e \psi
   & = & 
   \sum_{\gamma_i \in \Omega_{T+1}} \psi^*_{\gamma_1}
   a(\gamma_1(T),\sigma_1(T),\gamma_2(T),\sigma_2(T))\psi_{\gamma_2}  
\\ 
  & = & 
  \sum_{n_i,\sigma_i} a(n_1,\sigma_1,n_2,\sigma_2) 
  \left( \sum_{\gamma\in\Omega_{T-1}} 
      \psi_{\gamma \# n_1 \# (n_1+\sigma_1)}
  \right)^*
  \left( \sum_{\gamma'\in\Omega_{T-1}} \psi_{\gamma' \# n_2 \# (n_2+\sigma_2)}
  \right) \\ 
  & = & \sum_{n_i,\sigma_i} a(n_1,\sigma_1,n_2,\sigma_2) \rho^*_{n_1,
  \sigma_1} \rho_{n_2,\sigma_2} 
  \geq 0
  \ .
\ena
Here we have used 
a slightly different version of the $\#$ notation 
introduced after equation
(\ref{consistency}), 
according to which
$\gamma^{T+1}=\gamma^{T-1}\#\gamma^{T+1}(T)\#(\gamma^{T+1}(T)+\sigma(T))$.
In the final step of the proof, the inequality followed from the
positivity of $a$ as applied to the vector 
\be 
  \rho_{n,\sigma}
 =\sum_{\gamma\in\Omega_{T-1}} \psi_{\gamma \# n \# (n+\sigma)}
  \ .
\en 
The positivity of $D^e$ is proved in a similar way, but performing the
sums on $\sigma_i(T)$ first. 

This concludes the induction and the proof. 
Notice that the sum-rule (\ref{aconstr}) played no role in the proof and
$a$ need not have satisfied it.  Notice also that the restriction to a
nearest neighbor transfer matrix could have been lifted with little
difficulty.

We believe that the converse of the Theorem~\ref{Th1} is likely to
hold as well, at least generically; however we do not have a clean
proof of this.  If we are right, then, in seeking a transfer matrix
$a$ that will yield a strongly positive decoherence functional $D$, we
lose no generality in assuming that $a$ itself is a positive matrix.

In any case, we have established that to obtain a bi-additive, hermitian
symmetric, positive decoherence functional from a transfer matrix, it
suffices to find a positive transfer matrix satisfying (\ref{asym}).
Let us now concentrate on doing so.

\section{The transfer matrix and its Fourier transform} \label{transferM}
In the Introduction, we raised the question whether the quantal process
corresponding to the Schr{\"o}d\-ing\-er equation could arise as the
continuum limit of a quantal random walk.  Because of the delta function
at the truncation time $T$ in the Schr{\"o}dinger decoherence functional
(cf. equation (\ref{qmD})), one might try to write down a transfer
matrix incorporating an analogous $\delta$-function (or rather Kronecker
$\delta$), but this would not accomplish anything, because such a
$\delta$-function would immediately throw us back to the classical case
(\ref{Ex1}).

This highlights a subtle difference between the type of process we are
considering in this paper and the type of process considered in ordinary
unitary quantum mechanics (with or without added decoherence).  In our
scheme, the central object is the decoherence functional $D$ (or the
corresponding quantal measure $\mu$), and our transfer matrix $a$
evolves $D$ directly from one instant of time to the next.  The fact
that $D$ can be built up step by step in this manner is what we refer to
in terming our quantal processes ``Markovian''.\footnote
{From the standpoint of the quantal measure $\mu$, our scheme falls
 short of being fully Markovian, since it is the decoherence functional
 $D$ which evolves locally in time, and this evolution makes essential
 reference to extra information (the imaginary part of $D$) not
 contained in the measure itself.  
 (Of course, 
  because of interference, 
  even $\mu$ contains extra information not
  present in the measures of the individual trajectories, 
  and this furnishes a quite different reason why one might view quantal
  processes as not Markovian in an important physical sense,
  cf. the discussion of ``fuzzing'' in \cite{Sorqmti}.)}
A particular consequence is that one could define a ``density matrix''
$\rho(n_1,n_2,t)$ which would evolve from one time to the next via the
action of transfer matrix $a$ and which would record the value of $D$
on classes of paths ending at $n_1$ and $n_2$, respectively.

Calling such a $\rho$ a density matrix could be misleading, however,
since it does not correspond to what is normally called ``density
matrix'' in the context of unitary quantum mechanics.  Indeed, unitary
evolution (unlike the classical random walk) is {\it not} contained
within our general transfer matrix scheme as a special case; and no
special case of our quantal random walk could be unitary, {\it even if}
we dropped the bound on the step size.\footnote
{The same applies to extensions of unitary quantum mechanics which incoporate
``environmental dechoherence''}
The reason is that the
density matrix in the normal sense (call it $\rho'(n_1,n_2)$) does not
vanish when $n_1\not=n_2$, whereas the decoherence functional of unitary
quantum mechanics does,
because of the final $\delta$-function that occurs in (\ref{qmD}).  

Indeed, the normal density matrix $\rho'$ is related only indirectly to
the decoherence functional of unitary quantum mechanics and has no
meaning in itself, within a ``histories'' formulation.  Rather it is a
kind of phantom, telling us what the decoherence functional ``would be''
if one were not going to impose the final delta-function at some future
time $T$.  But of course, one does need to impose this condition in the
future (in the unitary context), and because of it, unitary evolution
could be seen as not truly Markovian.  For the same reason, unitary
evolution could also be seen as unnatural, in agreement with the
observation that the final $\delta$-function that couples $\gamma_1$ to
$\gamma_2$ is a rather violent and discontinuous interaction, compared
to what one might have expected for such a coupling.

By these comments, we do not mean to imply that unitary behavior cannot
arise from our type of process, but only that it cannot be obtained by
specialization.  (It could be at best some sort of singular limit, since
the transfer matrix, as we have employed that term, 
would be either infinite, zero, or undefined
if one tried to define it for the
case of unitary evolution.)  Rather, we would expect it to emerge only
after some suitable coarse-graining in both space and time.  This would
yield an effective decoherence functional on coarse-grained
trajectories, which would be effectively unitary if it took the form
(\ref{qmD}), in other words if the interaction between ``walker'' and
``antiwalker'' reduced to a final delta-function.  (Equivalently stated:
the amplitude should factor {\it except for} the final delta-function.)

But how could one design a transfer matrix that would lead to such a
result?  Could one begin with ``non-interacting'' paths $\gamma_1$ and
$\gamma_2$, each of which evolved approximately according to the
Schr{\"o}dinger equation, and then adjoin a suitably ``attractive''
interaction that at late times would condense into a smeared
delta-function?
The following developments are motivated by this prospect.  First, we
consider ``free Schr{\"o}dinger evolution'' and then ask how to
perturb it while preserving positivity.  In the end however, it proved
easier to find the most general positive transitive transfer matrix
directly, and we present this in Section 3.2.

\subsection{Transition amplitudes for the Schr{\"o}dinger process} 
\label{Schpart}
We are interested in finding a quantum random walk which could
describe, in the continuum limit, a particle satisfying the
Schr{\"o}dinger equation, 
\be
  \frac{\partial \psi}{\partial t}
  = \frac{i\hbar}{2m}\frac{\partial^2 \psi}{\partial x^2}.
\en
To this end, let us begin by approximating the Schr{\"o}dinger equation
via the method of finite differences, using the lattice,
$(m\Delta{t},n\Delta{x})$, $m\in\NaturalNumbers$, $n\in\Integers$.
We have, at leading order,
\bea
  \frac{\partial \psi(t,x)}{\partial t} 
  & = & 
  \frac{\psi(t +\Delta t,x) - \psi(t,x) }
  {\Delta t}\\
  \frac{\partial^2\psi(t,x)}{\partial x^2} & = & \frac{\psi(t,x+\Delta
  x)+\psi(t,x-\Delta x)-2\psi(t,x)}{(\Delta x)^2} \ ,
\ena
which yields to leading order
\bea
  \psi(t+\Delta t,x) & = & (1-2ip)\psi(t,x)+ip(\psi(t,x+\Delta x)+\psi(t,x-
  \Delta x)) \\
  \mbox{with } p & = & \frac{\hbar \Delta t}{2m(\Delta
  x)^2}. \label{scheq} 
\ena
{}From this expression, a set of one-particle transition amplitudes  
can be extracted as
\bea
  p(\pm 1) & = & ip\\
  p(0) & = & 1-2ip. 
\label{Stp} 
\ena 
These transition amplitudes sum to one like classical transition
probabilities, but of course they are neither real nor positive.

Let us denote by $p(\gamma)$ the amplitude of a truncated path obtained
from the Schr{\"o}dinger amplitudes $p(\pm1,0)$.
The quantum decoherence functional we are ultimately aiming for is 
something like
\be 
  D(\gamma_1^T,\gamma_2^T)=p^*(\gamma_1^T)p(\gamma_2^T)\delta_{
  \gamma_1^T(T),\gamma_2^T(T)} \ ,  
\label{DQ} 
\en
where $\delta$ is the Kronecker symbol.  However, as we know, this
decoherence functional with its delta-function interaction depending
only on the end-points of the truncated paths is not viable for a
nearest neighbor random walk, since it necessarily comes into conflict
with the additivity condition (\ref{bi-add}), which in this case is
equivalent to unitarity.  
(Nor, as we have already observed, would it be appropriate for a process
which could be called Markovian in our sense of the word.)
However (cf. (\ref{Ex2})), there is another decoherence functional which
can be derived from the transition amplitudes
$p(\gamma)$, given simply by 
\be
  D(\gamma_1^T,\gamma_2^T)=p^*(\gamma_1^T)p(\gamma_2^T)\ .
\label{nonID} 
\en
It would  not reproduce the Schr{\"o}dinger process in the continuum
limit, but it is Markovian, corresponding to the transfer matrix 
\be
  a^{(0)}(n_1,\sigma_1,n_2,\sigma_2)=p^*(\sigma_1)p(\sigma_2)\ , \label{pp}
\en 
which, so far, does not depend on the particle positions
$n_{1,2}$.  
Notice three things: first that the sum-rule (\ref{Qconstr}) is
automatically satisfied by (\ref{pp}) and doesn't need to be imposed
separately; second that $a^{(0)}$ is positive, as expected from the fact
that the decoherence functional (\ref{nonID}) from which it derives is
itself obviously strongly positive; and third that the symmetry
$p(+1)=p(-1)$ expresses a no--drift condition on our quantal walk.

This transfer matrix does not describe the quantum process we are
interested in.  However, if it is perturbed in such a way as to become
attractive, then at late times, the decoherence functional should vanish
when the two ``particles'' are far from one another, and this is exactly
the type of decoherence functional (\ref{DQ}) we are seeking.
Perturbing the transfer matrix while keeping it positive, however, is
far less trivial than it might seem, and this is the question we now turn
to.

\subsection{Positivity of the transfer matrix}
Henceforth, we will assume that the transfer matrix is defined for all
$n$ and $\sigma$, this being the generic case.  This implies in
particular that the consistency condition (\ref{Qconstr}) will have to
hold for all values of its arguments.

We saw in Eq. (\ref{aparam}) that the transfer matrix
depends on the two particle-positions only through the difference
$n_2-n_1$.  Using the reduced notation for the transfer matrix, we can
introduce the Fourier transform (of  period $2\pi$),
\be
 a(\theta,\sigma_1,\sigma_2)
 =\sum_{n=-\infty}^{+\infty} a(n,\sigma_1, \sigma_2) \e^{in\theta}
\en
and its inverse,
\be
  a(n,\sigma_1,\sigma_2) 
  =
  \frac{1}{2\pi}\int_0^{2\pi} a(\theta,\sigma_1,\sigma_2)
  \e^{-in\theta} d\theta .
\en 

Under Fourier transform, the hermiticity condition Eq. (\ref{asym})
becomes the requirement that the
$3\times3$ matrix $a(\theta,\sigma_1,\sigma_2)$ be hermitian for all
$\theta$.
The sum rule, equation (\ref{Qconstr}), can also be Fourier transformed
without difficulty, yielding:
\be
  \sum_{-1\leq \sigma_1,\sigma_2\leq 1}
  a(\theta,\sigma_1,\sigma_2)
  =2\pi\delta(\theta),  \label{IFconstr} 
\en
where $\delta$ is the usual delta distribution of Dirac.  Observe that
this equation just fixes the value of the quadratic form associated with
the hermitian matrix $a(\theta)$ on the vector $(1,1,1)$.

Now consider the requirement that $a$ be a positive matrix. 
Introducing a generic vector $\psi_{n,\sigma}$ and its associated
Fourier series 
\be
 \psi(\theta,\sigma)=\sum_{n=-\infty}^{+\infty}\psi_{n,\sigma}
 \e^{in\theta} \ , 
\en 
we obtain this condition in the form:
\bea 
\psi^\dag a \psi & = & \frac{1}{2\pi}\int_0^{2\pi} d\theta\, \sum_{n
_i,\sigma_i} a(\theta,\sigma_1,\sigma_2) \psi_{n_1,\sigma_1}^*
\psi_{n_2,\sigma_2} \e^{-i(n_2-n_1)\theta} \\  
& = & \frac{1}{2\pi} \int_0^{2\pi} d\theta \sum_{\sigma_i=-1}^1 \psi^*
_{\sigma_1}(\theta) \psi_{\sigma_2}(\theta)a(\theta,\sigma_1,\sigma_2)
\geq 0 \ .
\label{Ipos} 
\ena  
This is certainly satisfied for any $\psi$ if the hermitian $3\times3$
matrix $a(\theta,\sigma_1,\sigma_2)$ is positive for all $\theta$.
Conversely, if that matrix fails to be positive semi-definite at any
$\theta_0$, then there exists $\rho_i$ a three component vector such
that $\rho_i^\dag a(\theta_0,i,j)\rho_j<0$.  Then the vector 
\be
  \psi_{n,\sigma}=\frac{\rho_\sigma}{2\pi}\e^{-in\theta_0} 
\en
to which corresponds the Fourier transform 
$\psi(\theta,\sigma)=\rho_\sigma\delta(\theta-\theta_0)$,
will certainly yield $\psi^\dag a\psi<0$,
proving that $a$ is not positive.
Thus, we have demonstrated the following 
\begin{Theo} 
If the Fourier transform, $a(\theta,\sigma_1,\sigma_2)$,
of the interaction matrix $a(n,\sigma_1,\sigma_2)$ is a 
positive $3\times 3$ hermitian matrix, for all values of $\theta$,
then the corresponding transfer matrix is also positive,
and conversely.
\end{Theo}
Actually, our demonstration was not rigorous, mainly because $a(\theta)$
need not (and indeed cannot) be continuous.  Nonetheless, it seems
clear that the theorem does hold rigorously if $a(\theta)$ is
interpreted as a matrix-valued measure, because it is then nothing more
than the matrix generalization of the following classical
result.\footnote
{The condition in the Herglotz theorem that $f$ be ``of positive type'' means
 that $M(n_1,n_2):=f(n_1-n_2)$ be a positive matrix.  In other words, it
 is precisely the condition we need for positivity of the transfer
 matrix.  We have quoted the Herglotz theorem from \cite{herglotz}.}

\begin{Theo} ({\bf Herglotz})
  A function $f$ on the integers $\Integers$ is of positive type iff it
  is the Fourier transform of a finite (positive) measure on the circle
  $S^1$, i.e. $f$ is 
  of positive type iff it has the form
  $$f(n) = \int_{-\pi}^{\pi}  d{\alpha}(\theta)  \exp(i n \theta)$$
  where $\alpha$ is a bounded monotone increasing function on $[-\pi, \pi]$. 
\end{Theo}

The form $d\alpha(\theta)$ allows $a(\theta)$ to contain
delta-functions, but nothing more singular.  To see why this should be
so, consider a transfer matrix of the following form: 
\be
a(\theta,i,j)
=
a_c(\theta,i,j)+\sum_{k=0}^{l} M^{(k)}_{ij}\delta^{(k)}(\theta-\theta_0), 
\label{dexp} 
\en
where $a_c(\theta,i,j)$ is a positive continuous (or even piecewise continuous) 
contribution, $\theta_0$ is a fixed number, $\delta^{(k)}$ is
the $k$-th derivative of the Dirac distribution and $l$ is a given
integer, for which $M_{ij}^{(l)}\not=0$. 
Let us prove then that the positivity condition implies that $l=0$,
i.e. that nothing more singular than a $\delta$-function can be
present, 
and that $M^{(0)}_{ij}$ must be a positive hermitian matrix.

To that end, let us suppose that $l>0$ and introduce smooth test functions
$\psi_l^\pm (\theta)$ with a support on the interval $[-1,1]$ and such
that 
\be 
 \left. \frac{d^k\psi_l^\pm}{d\theta ^k}\right| _0
 =
 \delta_{0,k} \pm \delta_{l,k},\   0\leq k \leq l 
\en
where $\delta$ is the Kronecker symbol. 
Then taking
$\psi_\sigma (\theta)=  \rho_\sigma \psi_l^\pm (\alpha
(\theta-\theta_0))$ with $\alpha>0$ a real parameter, the inequality
(\ref{Ipos}) yields 
\be  
\psi^\dag a \psi = \frac{1}{2\pi}\rho_i^\dag \rho_j 
(\int_0^{2\pi}d\theta |\psi_l^\pm (\alpha (\theta-\theta_0))|^2
a_c(\theta,i,j)+M^{(0)}_{ij}\pm 2\alpha^l M^{(l)}_{ij}) \geq 0 .\en 
Bounding from above the first term on the left--hand side then yields \be
\rho_i^\dag \rho_j (\min(\frac{2}{\alpha},2\pi) \sup_{0\leq \theta
  \leq 2\pi} [ \, |\psi_l^\pm (\alpha (\theta-\theta_0))|^2 a_c
(\theta,i,j)] +M^{(0)}_{ij}\pm 2\alpha^l M^{(l)}_{ij}) \geq
0.\label{l0ineq}\en 
In the large $\alpha$ limit, the linear term, which dominates
the left--hand side, can only be
positive if it is identically zero, thus yielding that $\rho_i^\dag
M^{(l)}_{ij} \rho_j=0$ for any vector $\rho$, which in turn implies
that $M^{(l)}_{ij}=0$ in contradiction with its definition. Thus $l=0$.
Furthermore, with $l=0$, the inequality (\ref{l0ineq}) becomes \be
\rho_i^\dag \rho_j (\min(\frac{2}{\alpha},2\pi) \sup_{0\leq \theta
  \leq 2\pi} [ \, |\psi_l^\pm (\alpha (\theta-\theta_0))|^2 a_c
(\theta,i,j)] +M^{(0)}_{ij}) \geq 0, \en
which in the limit $\alpha\rightarrow +\infty$ yields immediately that
$M^{(0)}_{ij}$ is a positive matrix as desired.

It should also be clear that this proof can be generalized without any
difficulty to include a finite number of delta functions at different
points, and for convenience, we will not consider anything more general
than this. Thus, we have the  
\begin{Corol} 
The Fourier transform of a distribution of the form 
\be 
  a(\theta,i,j)=a_c(\theta,i,j)+\sum_{k=0}^{L} a^{(k)}_{ij}
  \delta(\theta-\theta_k),
  \label{Icdel} 
\en 
where $a_c(\theta,i,j)$ is continuous 
(or piecewise continuous),
hermitian and positive for all $\theta$, and where the $a^{(k)}_{ij}$
are positive hermitian matrices, is a positive transfer matrix. 
\end{Corol}
Strictly speaking, we do not have a transfer matrix until
the sum rule (\ref{Qconstr}) is also satisfied.  The conditions needed
to insure this are implemented explicitly in the next section. 

Although the form (\ref{Icdel}) is not quite exhaustive, it furnishes a
large class of positive transfer matrices, and therefore a large class
of strongly positive decoherence functionals.

\section{Parameterizing the positive transfer matrices} \label{pptm}
Before turning to the general case, let us briefly treat the classical
random walk, which is a special case of our Markovian process, as we
have seen.  In order to be able to apply the above Corollary without
modification, we need the transfer matrix to be defined for all possible
arguments, $n$ and $\sigma$, and we may choose to accomplish this by 
setting it to 0
outside its natural domain of definition.  It is clear that
it will be positive after this extension if and only if it was positive
before the extension.  
When defined in this way, it remains diagonal and independent
of $n$, taking specifically the following form:
\be
 a(n_1,\sigma_1,n_2,\sigma_2)
 =
 p_\sigma \delta_{\sigma\sigma_1} \delta_{\sigma \sigma_2} \ , 
\label{aclas} 
\en
where $\delta$ is the Kronecker symbol.  
In the absence of any position
dependence, the Fourier transform is trivial and yields
$a(\theta,\sigma_1,\sigma_2)
 =2\pi p_\sigma\delta_{\sigma\sigma_1}\delta_{\sigma\sigma_2}\delta(\theta)$.
The corollary then tells us that $p_\sigma\geq 0$,
where, 
to satisfy the sum rule (\ref{Qconstr}), or equivalently in
this case (\ref{clasconstr}), 
we need also $\sum_\sigma p_\sigma =1$.  

Thus we recover exactly the expected form for a classical random walk.
For an isotropic walk, the rightward and leftward probabilities must
coincide, and the parameters $p(\sigma)$ of the process simplify to just
$(p,1-2p,p)$, with a single degree of freedom $p\in [0,1/2]$.

Now let us turn to the generic case, whose sum rule is given by equation
(\ref{Qconstr}).

\subsection{The generic case} \label{TGQ}
At this point, it is convenient to split the Fourier transformed function
$a(\theta,\sigma_1,\sigma_2)$ of (\ref{Icdel}) into a fixed
part $a^{(0)}$ chosen as simply as possible to satisfy the Fourier
transform of (\ref{Qconstr}), 
namely
\be 
  \sum_{\sigma_1,\sigma_2} a^{(0)}(\theta,\sigma_1,\sigma_2)=2\pi
  \delta(\theta)\ ,
\label{afconstr} 
\en
and a variable part $a_i(\theta,\sigma_1,\sigma_2)$ which satisfies the
homogeneous version of the same equation:
\be
  \sum_{-1\leq\sigma_1,\sigma_2\leq 1}a_i(\theta,\sigma_1,\sigma_2)=0\ .
\label{iconstr}
\en
Thus, we shall write the general solution in the form 
\be
 a(\theta,\sigma_1,\sigma_2)
 =
 2\pi a^{(0)}_{\sigma_1\sigma_2}\delta(\theta) +
 a_i(\theta,\sigma_1,\sigma_2)
\label{agen} \ ,
\en
where $a_i$ can be split further into a continuous part and
a distributional part,
\be 
 a_i(\theta,\sigma_1,\sigma_2)
 =
 a_c(\theta,\sigma_1,\sigma_2)
 +
 \sum_{k=1}^{L} a^{(k)}_{\sigma_1\sigma_2} \delta(\theta-\theta_k) \ ,
\label{cdsplit} 
\en
where the $\theta_k$ for $k=1\cdots L$ are any distinct nonzero angles.
We have seen that, in order to generate a proper decoherence functional,
all the $3\times3$ matrices $a^{(j)}$ must be hermitian positive, as must
also the matrices $a_c(\theta,\sigma_1,\sigma_2)$ for each $\theta$.
Furthermore, these matrices must satisfy the constraint equations
\bea
 \sum_{-1\leq \sigma_1,\sigma_2\leq 1} a^{(0)}_{\sigma_1,\sigma_2} 
 & = & 1 \\
\sum_{-1\leq \sigma_1,\sigma_2\leq 1}  a_c(\theta,\sigma_1,\sigma_2)=
\sum_{-1\leq \sigma_1,\sigma_2\leq 1} a^{(k)}_{\sigma_1\sigma_2}  & =
& 0 \ .  \label{vconstr} \ena 

Simple choices for the matrix $a^{(0)}$ are easy to find, for instance
$a^{(0)}_{ij}=1/9$.  In practice, the choice of $a^{(0)}$ would be
adapted to the problem under consideration.  For instance, as discussed
in section 3, it would be natural to expect the transfer matrix whose
coarse-grained evolution reproduces the Schr{\"o}dinger process to be a
perturbation of the transfer matrix given in (\ref{pp}).  Thus, one
might use the latter for $a^{(0)}$, since it fulfills all the
requirements, as observed earlier.

As for the variable part $a_i$, it must satisfy the homogeneous
constraint (\ref{iconstr}).  We can simplify this condition by observing
that it merely signifies that the vector $(1,1,1)$ is in the kernel of
the positive quadratic form associated with the hermitian matrix
$a_i(\theta)$.  This in turn implies that $(1,1,1)$ is a minimum of the
quadratic form, and therefore an eigenvector of $a_i(\theta)$ for the
eigenvalue zero.  Conversely, if $(1,1,1)$ is a zero eigenvector of
$a_i$ then the latter obviously satisfies the constraint.
To summarize, the constraint
(\ref{iconstr}) is equivalent to the three equations 
\be 
\sum_{-1\leq \sigma\leq 1} a_i (\theta,\tau,\sigma)=0 \ ,
\en
or equivalently,
\be
 \sum_{-1\leq \sigma\leq 1} a_c(\theta,\tau,\sigma)=\sum_{-1\leq
   \sigma\leq 1} a^{(k)}(\tau,\sigma) =0\label{kerc}  \en
for all $\tau \in \braces{-1,0,1}$.  

It follows that both these matrices must have the form 
$0_{(1,1,1)}\oplus H_{(1,1,1)^\perp}$, 
where $H$ is an arbitrary positive hermitian
matrix on the two--dimensional space $(1,1,1)^\perp$.
Now, the general two--dimensional positive hermitian matrix can be
parametrized as 
\be 
\left( \begin{array}{cc} A & \sqrt3 (R-iI) \\ \sqrt3 (R+iI) & 3B
\end{array}\right), 
\en 
where $A,B,R,I$ are real numbers
subject to the inequalities
\be
 A\geq 0,\ AB\geq(R^2+I^2) \ .    \label{ABCconstr} 
\en

Then, employing
$((1/\sqrt2,0,-1/\sqrt2),(1/\sqrt3,-2/\sqrt3,1/\sqrt3))$
as an orthonormal basis for $(1,1,1)^\perp$, one straightforwardly
derives the general solution for the matrices $a_c$ and $a^{(k)}$ as 
\be
 \left( 
 \begin{array}{ccc} \frac{A+B}{2}+R &
 -B-R+iI & \frac{B-A}{2}-iI \\ -B-R-iI & 2B & -B+R+iI \\ \frac{B-A}{2}+
 iI & -B+R-iI & \frac{A+B}{2}-R 
\end{array} 
\right)
\label{genherm} \ ,
\en
still subject to (\ref{ABCconstr}).

In the following, we will further require that the reflection symmetry
of the non--interacting model be preserved by the interaction term,
i.e. that there be no drift.  This implies that the transfer matrix be
symmetric under the reflection $n\rightarrow -n$:
\be
 a(-n_1,-\sigma_1,-n_2,-\sigma_2)=a(n_1,\sigma_1,n_2,\sigma_2)
 \label{adrift}  \ .
\en 
Under Fourier transformation, this becomes 
\be 
  a(\theta,\sigma_1,\sigma_2)=a(-\theta,-\sigma_1,-\sigma_2) \ ,
\label{ateq}
\en
which on substituting into Eq. (\ref{agen}) yields the new general form 
\bea
 a(\theta,\sigma_1,\sigma_2)
 & = &
 2\pi\delta(\theta) a^{(0)}_{\sigma_1\sigma_2}
 +a_i(\theta, \sigma_1,\sigma_2) \label{agen2a} \\
 a_i(\theta, \sigma_1,\sigma_2) & = & a_c(\theta, \sigma_1,\sigma_2)
 +
 \sum_{k=1}^{L} (a^{(k)}_{\sigma_1 \sigma_2}  \delta(\theta-\theta_k)
               + a^{(k)}_{-\sigma_1 -\sigma_2}\delta(\theta+\theta_k))
 \ ,
 \label{agen2} 
\ena
where $a^{(k)}$ and $a_c$ are of the general form
(\ref{genherm}), with $a^{(0)}$ and $a_c$ satisfying as well the
additional condition (\ref{ateq}).  In the case of 
$a_c$, this says simply that $A$ and $B$ are even
functions of $\theta$, while $R$ and $I$ are odd.
(Positivity places no extra
constraint on $a^{(k)}$ since $a^{(k)}_{-i -j}$ is automatically
positive if $a^{(k)}_{ij}$ is.)

For the constant component $a^{(0)}$ of the transfer matrix,
a straightforward generalization of Eq.(\ref{pp}), which we
derived for the Schr{\"o}dinger process, provides
immediately a family of options: 
\be 
a^{(0)}(\sigma_1,\sigma_2)=p^*(\sigma_1)p(\sigma_2) \ ,
\en
with $p(-1)=p(1)=p$, $p(0)=1-2p$, and $p$ any complex number. 
In this
case, the variable part  $a_i$ of the transfer matrix can be identified
with the interaction term discussed in Section \ref{Schpart}.

In conclusion, we have the following theorem: 
\begin{Theo}
  Given a fixed positive hermitian matrix $a^{(0)}$ whose components sum
  to unity, any transfer matrix whose Fourier transform has
  the form (\ref{agen}), (\ref{cdsplit}), (\ref{genherm}), subject to the
  inequalities (\ref{ABCconstr}), gives rise to a strongly positive
  decoherence functional.  
  Furthermore, if the Fourier transform has
  the more restrictive form 
  (\ref{agen2a}), (\ref{agen2}), 
  with $a^{(0)}(\sigma_1,\sigma_2)=a^{(0)}(-\sigma_1,-\sigma_2)$
  and with
  $A$ and $B$ even and
  $R$ and $I$ odd in $a_c$, then the corresponding quantal random walk is
  without drift.
\end{Theo} 

\subsection{A simple example}
Since we introduced the example of the ``non-interacting Schr{\"o}dinger
process'' in subsection \ref{Schpart} and argued that the type of
quantal random walk of greatest interest would be a perturbation of it,
we will use such a perturbation to illustrate the general solution just
found. 

Thus, we take $a^{(0)}$ to be given by Eq. (\ref{pp}),
where $p$ is the parameter introduced in Eq. (\ref{scheq}). 
For simplicity, let us choose a solution without drift of the
type (\ref{agen2}), with $a_c=0$, $L=1$, and 
\be
a^{(1)}(i,j)
=
\pi p^2\left(
   \begin{array}{ccc} 1 & 0 & -1 \\ 0 & 0 & 0 \\ -1 & 0 &  1 \end{array}
\right) \ .
\en
The resulting transfer matrix, 
\be
  a(n_1,\sigma_1,n_2,\sigma_2)=
 \left(\begin{array}{ccc} p^2[1+\cos((n_2-n_1)\theta_1)] &  -p(2p+i) & 
p^2[1-\cos((n_2-n_1)\theta_1)] \\ -p(2p-i) & 1+4p^2 & -p(2p-i) \\ p^2
[1-\cos((n_2-n_1)\theta_1)] & -p(2p+i) & p^2[1+\cos((n_2-n_1)\theta_1)]
\end{array}\right) 
\en 
defines a strongly positive decoherence functional and a quantal measure through
Eq. (\ref{Ddef}). 
It is particularly simple for $\theta_1=\pi$, being 
periodic with period $2$ in $n_2-n_1$.

\section{Conclusion} \label{conclusion}

It is noteworthy that one is led to go beyond unitary quantum
mechanics in order to arrive at a quantal analog of the random walk.  In
the wider framework we have implemented in this paper (that of ``quantum
measure theory'' or ``generalized quantum theory'') unitarity is
relinquished but a positivity requirement is retained, as described in
detail above.  The resulting dynamics furnishes an interesting example
of a quantal process belonging to this more general framework.

The family of transfer matrices $a$ derived in Sections \ref{transferM}
and \ref{pptm} is essentially the most general possible in this
framework, for a discrete-time, homogeneous and isotropic, random walk
on the integers with nearest neighbor transitions and no memory.  In
addition to these conditions, the only further input to our derivation
was the requirement that the transfer matrix $a$ be positive.  As we
saw, this implies strong positivity of the decoherence functional $D$
and {\it a fortiori} positivity of the quantum measure $\mu$, i.e. the
inequality (\ref{posi}).

We do not believe there is much room for weakening these positivity
conditions.  Since  positivity of the transfer matrix is 
sufficient and almost certainly necessary 
for strong positivity of $D$, the only real question is
whether one might consider weakening the latter to weak
positivity (\ref{posi}); but it is not clear
how much more generality that would achieve, while on the other hand
there are good independent reasons for imposing strong positivity in 
its own right
(two such being the consequent existence of a Hilbert space and the
 guarantee that mentally combining two non-interacting subsystems
 into a single system will not ruin positivity\footnote%
 {In this way, strong positivity is the analog of what, in another
  context, is called ``complete positivity'' for the evolution in time
  of a quantum mechanical density
  operator \cite{complete-positivity}.}).

As we have seen, positivity has proven to be remarkably restrictive, and
we have been able to find the most general possible transfer matrix that
satisfies it.  Nevertheless, the resulting random walk still has more
free parameters than are present classically, where the homogeneous,
isotropic walks with step size less than or equal to 1 can be
parameterized by a single real number, as in Eq. (\ref{aclas}).  In
contrast, our general transfer matrix contains two even and two odd real
{\em functions}.

An obviously important question is whether some instance of our scheme
admits a unitary continuum limit\footnote
{This question is also very relevant for quantum gravity, for
 reasons given in the Introduction.}.
To a large degree, this seems to be equivalent to asking 
whether the ``interaction'' between the paths $\gamma_1$ and $\gamma_2$ 
can be chosen so as to reduce in the limit 
to a final delta function,
as in equation (\ref{qmD}) or (\ref{DQ}); 
for such a delta-function interaction, 
were it exact, 
would immediately imply unitarity of the transfer matrix.  
If one can come up with an interaction term that does this, 
then as a consequence 
a quantal ``law of large numbers'' should take effect, 
or at least this seems very plausible.  
In fact, 
the Markov property should produce 
in the continuum limit 
a differential equation of first order in time, while 
the locality of the walk (bounded step size) should also imply 
a differential equation in space.   
On dimensional grounds\footnote
{Or, if you like, arguing from a gradient expansion, as is done in the
 context of the ``renormalization group''.},
the surviving terms in the continuum limit would then be those of lowest
possible differential order, namely those of the Schr{\"o}dinger
equation.  This would be in close analogy to how the diffusion equation
emerges as the continuum limit of the random walk in the classical
setting.

To get this kind of delta function behavior one would want the
interaction term $a_i$ in (\ref{agen2}) 
to be suitably ``attractive''.
The parameterization we adopted for the transfer matrix in
Section~\ref{TGQ} was chosen with this possibility of a
Schr{\"o}dinger limit in mind, and specifically with the hope that the
desired interaction terms would appear as mild perturbations in this
parameterization.  (Logically speaking however, if there is in fact a
``quantal central limit theorem'', it would be enough to find any
interaction that reduced to a $\delta$-function, using any
parameterization.)  Based on a preliminary investigation of these
issues, we believe the main difficulty in finding an interaction of
the desired sort is tied up with the positivity conditions.  As we
have seen, the ``Markov sum rule'' (\ref{Qconstr}) implies that the
`interaction part' of the transfer matrix has a nonzero kernel.  But
this means one is always on the verge of ruining positivity, and one is
not free to introduce whatever interaction one wishes.  We hope to
return to these questions in a future work.

It seems appropriate to conclude this paper with some comments on the
physical interpretation of the quantal processes we have defined herein.
Obviously, we cannot say anything definitive here because the
interpretation of quantum mechanics itself is still not settled.
Indeed, the concepts of decoherence functional and quantal measure were
introduced in part in the hope of making progress on precisely these
interpretational issues by avoiding the invocation of undefined
`measurements' carried out by external agents inhabiting an external,
classical world or ``laboratory''.

At least two interpretive schemes have emerged from these efforts,
neither of which is completely worked out yet.  Both schemes seek to
endow the quantal measure $\mu$ with an {\it intrinsic} meaning that
does not need to refer to any external agents at all.  The first, and
presently more complete scheme aims in effect to realize Bohr's
``classical world'' within the quantal world by identifying it with a
set of decohering variables \cite{gen,griffiths}.  Within this scheme,
$\mu$ would be interpreted as an ordinary probability, but only after
restriction to a suitable subset $\R'\subset\R$, namely a subset for
which this interpretation would be consistent because all the
interference terms $I(A,B)$ would vanish between members of $\R'$
(``decoherence'').  Adapting this scheme to our quantal random walk, one
might, for example, take for $A$ the set of all paths which ever return
to the origin and ask whether $A$ decoheres with its complement (call it
$B$).  If it did, then one could potentially interpret $\mu(A)$ and
$\mu(B)=1-\mu(A)$ as probabilities in the normal sense.  (Incidentally,
if $\mu(B)=0$, as one might expect, then $I(A,B)=0$ would follow
immediately from strong positivity and Theorem \ref{prodineq}.)

The second intrinsic scheme does not demand decoherence and does not
attempt to equate $\mu$ to a classical probability.  Instead, it seeks
inspiration from a different feature of the measurement process as it is
usually conceived, namely the establishment of a correlation between
subsystem and apparatus.  It aims to formulate this notion of
correlation intrinsically, and it takes such correlations to be
objective features of the world which are predictable in terms of the
measure $\mu$ \cite{Sorqmti}.  Clearly, this scheme could not apply to
our random walk {\it per se}, because there is nothing for the walker to
correlate with.  One would need to add something like a second walker,
or at a minimum some finite state ``meter'' which could correlate with
some attribute of the walker's trajectory.  Beginning with the simplest
possibility, one might seek to introduce a second subsystem with trivial
internal dynamics, but coupled momentarily to the walker in such a way
as to become correlated with some position variable of interest.  The
enlarged system would again be described by a quantal measure 
$\mu'$ and decoherence functional $D'$, and one could again require
strong positivity of the latter.  The question then would be under what
circumstances such a coupling was possible.  (Notice that the original
measure $\mu$ would play its role indirectly here, by influencing the
possible couplings, consistent with strong positivity.)  An intriguing
question of this sort is whether one could design a ``did it ever
return?'' meter in this manner.

Of course, one could also consider the random walks derived herein more
along traditional lines, more in the spirit of the ``Copenhagen
interpretation'', which {\it does} invoke external observers or
instruments to which the rules of classical probability theory are
assumed to apply simply by fiat.  The nature of such observers is
inherently undefined, of course, but the ``von Neumann scheme'' permits
them to be modeled as quantum systems, at the cost of having to
introduce some still more primitive external agents to measure the
measurers.  Because it is logically circular, this process of ``moving
the cut'' does not really explain the measurement process, but it does
establish a certain self consistency of the textbook measurement
paradigm.  One could try to formulate an analogous ``measurement
theory'' for generalized quantum mechanics (in terms of $D$ and $\mu$
rather than the state-vector) and apply it to our random walk.  This
could mean figuring out how to couple an effectively classical
``apparatus'' to, e.g., the walker's position at a certain time.  Since
different positions of the walker will not in general decohere in our
case, no classical apparatus could be expected to reproduce the quantal
measure of a ``position predicate'' directly as the probability of a
meter reading.  (That is why the ``consistent historians'' demand
decoherence in the first place, of course.)  Hence, to make the scheme
go through in its familiar form, one would have to find other variables
that did in fact decohere, and therefore ``could be measured''.

Alternatively, one might seek a ``self consistent'' measurement model
that invoked external agents, but did not try to reduce quantal measures
to classical ones (probabilities).  In other words, one would allow
``meter readings'' to interfere, and would seek to couple the quantal
walker to a quantal meter in such a way that the quantal measure of a
given meter reading after coupling would coincide with the quantal
measure of the corresponding set of walker trajectories in the uncoupled
case.  This would be more completely parallel to the usual paradigm in
the sense that it would exhibit the self consistency of this sort of
``measurement'', without shedding any further light on the meaning of
$D$ or $\mu$ as such.

A rather different way to seek an interpretation of our quantal random
walk would be to ask whether one could {\it realize} it using ordinary
unitary processes (see \cite{knight_etal} for a related discussion).
Could one, for example, simulate it on a computer?  We suspect that this
would be impossible with a classical computer, but it might be possible
with a quantum one.  Since our random walk is not unitary, it could only
be realized, within a unitary framework, as an open system (or more
generally an incomplete set of variables).  Could one, then, program a
quantum computer so that its decoherence functional would reproduce that
of the random walk when restricted to a suitable set of discrete
variables?  (In addressing this question, the affinity between strong
and complete positivity might be helpful, since it is known that any
completely positive dynamics can be reproduced as the effective dynamics
of a subsystem of a larger unitary system.)

Thus, there remain many open questions for the future, whose
investigation promises to yield further insight into the quantum world.

\bigskip
\noindent
{\bf Acknowledgments:} We wish to thank 
A.P. Balachandran, 
Fay Dowker,
Raquel S. Garc{\'\i}a,
Joe Henson,
Giorgio Immirzi,
and
Siddhartha Sen
for their comments.
This work has been supported by 
CONACyT under grants E120.0462 and 30422-E, by
NSF under grants INT-0203760 and PHY-0098488, 
and by a European Community Marie Curie Fellowship.
The work also received partial support from the
EU Research Training Network in Quantum Spaces-Noncommutative Geometry QSNG.


\begin{thebibliography}{xx}

\bibitem{SpaceCS} 
L.~Bombelli, J.~Lee, D.~Meyer and R.D.~Sorkin, 
``Spacetime as a Causal Set'', 
  \journaldata {Phys. Rev. Lett.}{59}{521-524}{1987};
\hfil\break
Rafael D.~Sorkin,
``Causal Sets: Discrete Gravity (Notes for the Valdivia Summer School)'',
in the proceedings of the Valdivia Summer School, 
held January 2002 in Valdivia, Chile, 
edited by Andr{\'e}s Gomberoff and Don Marolf 
(to appear);
\eprint{gr-qc/0309009}
\hfil\break
David D.~Reid, ``Discrete Quantum Gravity and Causal Sets'',
\journaldata{Canadian Journal of Physics}{79}{1-16}{2001}
 \eprint{gr-qc/9909075}.


\bibitem{lambdapred}
R.D.~Sorkin, 
``First Steps with Causal Sets'', 
  in R. Cianci, R. de Ritis, M. Francaviglia, G. Marmo, C. Rubano, 
     P. Scudellaro (eds.), 
  {\it General Relativity and Gravitational Physics,} 
   (Proceedings of the Ninth Italian Conference of the same name, 
     held Capri, Italy, September, 1990), pp. 68-90
  (World Scientific, Singapore, 1991); \hfil\break
R.D.~Sorkin,
``Forks in the Road, on the Way to Quantum Gravity'', talk 
   given at the conference entitled ``Directions in General Relativity'',
   held at College Park, Maryland, May, 1993,
   {\it Int. J. Th. Phys.} {\bf 36}: 2759--2781 (1997)   
[arXiv:gr-qc/9706002]; \hfil\break
R.D.~Sorkin, 
``Discrete Gravity'',
a series of lectures to the 
{\it First Workshop on Mathematical Physics and Gravitation},
 held Oaxtepec, Mexico, Dec. 1995
 (unpublished).

\bibitem{rideout} D. Rideout and R. Sorkin ``A Classical Sequential
Growth Dynamics for Causal Sets'' {\it Phys. Rev.} {\bf D61} 024002
(2000), [arXiv:gr-qc/9904062].

\bibitem{mors} X. Martin, D. O'Connor, D. Rideout and R. D. Sorkin,
``On the ``renormalisation'' transformations induced by cycles of
expansion and contraction in causal set cosmology'', {\it Phys. Rev.}
{\bf D63}, 084006 (2001) and 
[arxiv:gr-qc/0009063].

\bibitem{avner}
Avner Ash and Patrick McDonald,
``Moment problems and the causal set approach to quantum gravity'',
\journaldata{J. Math. Phys.}{44}{1666-1678}{2003}
[arXiv:gr-qc/0209020].

\bibitem{anpa} 
{Graham Brightwell}, {H. Fay Dowker}, {Joe Henson} and {Rafael D.~Sorkin},
``General Covariance and the `Problem of Time' in a Discrete Cosmology'',
 in K.G.~Bowden, Ed.,   
 {\it Correlations}, Proceedings of the ANPA 23 conference,
 held August 16-21, 2001, Cambridge, England 
 (Alternative Natural Philosophy Association, London, 2002), pp 1-17
[arXiv:gr-qc/0202097]

\bibitem{observables}
Graham Brightwell, Fay Dowker, Raquel S.~Garc{\'\i}a, Joe Henson and 
     Rafael D.~Sorkin,
``{}``Observables'' in Causal Set Cosmology'',
\journaldata{Phys. Rev. D}{67} {084031} {2003},
[arXiv:gr-qc/0210061].

\bibitem{dam} 
D.~A.~Meyer, 
``On the Absence of Homogeneous Scalar Unitary Cellular Automata'',
\journaldata{Phys. Lett. A}{223}{337--340}{1996},
[arXiv:lquant-ph/9604011];
D.~A.~Meyer, 
``From Quantum Cellular Automata to Quantum Lattice Gases''
\journaldata{J. Stat. Phys.} {85} {551} {1996};
Todd.~A.~Brun,
``Quantum Walks driven by many coins'', 
[arxiv:quant-ph/0210161];
%
%
Todd A. Brun, Hilary A. Carteret and Andris Ambainis.
``The quantum to classical transition for random walks'',
\journaldata{Phys. Rev. Lett.} {91}{130602-1} {2002},
[arXiv:quant-ph/0208195].

\bibitem{bogh}
B.M.~Boghosian and W.~Taylor IV, "A quantum lattice-gas model for the
  many-particle Schr{\"o}dinger equation in $d$ dimensions",
  quant-ph/9604035
  
\bibitem{lambdamodel} 
Maqbool Ahmed, Scott Dodelson, Patrick Greene and Rafael D.~Sorkin,
``Everpresent $\Lambda$'',
[arXiv:astro-ph/0209274].

\bibitem{gen} J. Hartle ``Spacetime Quantum Mechanics and the quantum
Mechanics of Spacetime'', in {\it Les Houches, session LVII, 1992,
Gravitation and quantisations} (Elsevier Science B.V. 1995),
[arXiv:gr-qc/9304006]. 

\bibitem{Sorqmt} R. Sorkin ``Quantum Mechanics as Quantum measure
Theory'', {\it Mod. Phy. Lett.}, {\bf A9}, 3119--3127 (1994),
[arXiv:gr-qc/9401003].

\bibitem{Sorqmti} R. Sorkin ``Quantum measure Theory and its
interpretation'', in {\it Quantum Classical Correspondence:
Proceedings of the 4th Drexel Symposium on Quantum Nonintegrability},
229--251 (International Press, Cambridge Mass. 1997),
[arXiv:gr-qc/9507057]. 

\bibitem{rob}
Roberto B.~Salgado, ``Some Identities for the Quantum Measure and its 
 Generalizations'',
 \journaldata{ Mod. Phys. Lett.} { A17}{711-728}{2002},
[arXiv:gr-qc/9903015]

\bibitem{Chryssomalakos:2003ms}
C.~Chryssomalakos and M.~Durdevich,
``Higher Order Measures, Generalized Quantum Mechanics and Hopf Algebras'',
[arXiv:quant-ph/0309092].


\bibitem{Sorkin_role_oftime} R. Sorkin, ``On the Role of Time in the
Sum over Histories Framework for  Gravity'', {\it Int. J. Theor. Phys.} 
{\bf 33}, 523-534 (1994).

\bibitem{Glimm:1981xz}
J.~Glimm,  and A.~Jaffe,
{\it "Quantum Physics. A Functional Integral Point of View''},
New York, Usa: Springer ( 1981).


\bibitem{herglotz}
{\it Encyclopedic Dictionary of Mathematics, Second Edition} 
(MIT press, 1993), 
Article 192 B: ``Harmonic Analysis''

\bibitem{complete-positivity}
G\"oran Lindblad,
``Completely Positive Maps and Entropy Inequalities''
 \journaldata{Commun. Math. Phys.}{40}{147-151}{1975}


\bibitem{griffiths}
 R.~Griffiths, \journaldata{J.~Stat. Phys.}{36}{219}{1984};
 R.~Omnes, \journaldata{Rev. Mod. Phys} {64}{339}{1992}.

\bibitem{knight_etal}
Peter~L. Knight, Eugenio~Rold\'an and J.E. Sipe,
``Optical Cavity Implementations of the Quantum Walk''
[arXiv:quatu-ph/0305165].

\end{thebibliography}
\end{document}